\documentclass[useAMS,usenatbib,usegraphicx]{mn2e} 
\title[NGC~1407 and NGC~1400: star formation and chemical evolution]
{The early-type galaxies NGC~1407 and NGC~1400 $-$~II:
star formation and chemical evolutionary history} 

\author[M. Spolaor et al.]{Max Spolaor$^1$\thanks{E-mail: mspolaor@astro.swin.edu.au}, 
Duncan A. Forbes$^1$, Robert N. Proctor$^1$, George K. T. Hau$^{1,2}$, \newauthor
and Sarah Brough$^1$\\ 
$^1$Centre for Astrophysics \& Supercomputing, Swinburne University, Hawthorn, VIC 3122, Australia\\ 
$^2$Department of Physics, University of Durham, South Road, Durham, DH1 3LE, UK}

\begin{document}
\date{Accepted... Received...; in original form 2007}

\pagerange{\pageref{firstpage}--\pageref{lastpage}} \pubyear{2007}

\maketitle
\label{firstpage}

\begin{abstract}
We present a possible star formation and chemical evolutionary history for two early-type galaxies NGC~1407 and NGC~1400. They are the two brightest galaxies of the NGC~1407 (or Eridanus-A) group, one of the 60 groups studied as part of the Group Evolution Multi-wavelength Study (GEMS).

Our analysis is based on new high signal-to-noise spatially resolved integrated spectra obtained at the ESO 3.6m telescope, out to $\sim$~0.6 (NGC~1407) and $\sim$~1.3 (NGC~1400) effective radii. Using Lick/IDS indices we estimate luminosity-weighted ages, metallicities and $\alpha$-element abundance ratios. Colour radial distributions from HST/ACS and Subaru Suprime-Cam multi-band wide-field imaging are compared to colours predicted from spectroscopically determinated ages and metallicities using single stellar population models. The galaxies formed over half of their mass in a single short-lived burst of star formation ($\geq$~100~M$_{\sun}$/year) at redshift z~$\geq$~5. This likely involved an outside-in mechanism with supernova-driven galactic winds, as suggested by the flatness of the $\alpha$-element radial profiles and the strong negative metallicity gradients. Our results support the predictions of the revised version of the monolithic collapse model for galaxy formation and evolution. We speculate that, since formation the galaxies have evolved quiescently and that we are witnessing the first infall of NGC~1400 in the group.
\end{abstract}


\section{INTRODUCTION}
The physical formation and evolution processes taking place at various locations within a galaxy leave their fossil imprint in the stars. These chemodynamical imprints are the regular features observed in the stellar populations, photometry, internal kinematics and structure of early-type galaxies. Furthermore, these features are found to correlate defining scaling relations of remarkable tightness, such as the colour-magnitude relation and the fundamental plane. This class of galaxy, therefore, is one of the best guides to compare observations and theoretical model predictions. 

Theoretical scenarios that describe the formation and evolution of early-type galaxies include: a revised version of monolithic collapse, the dissipative (wet/gas-rich) and dissipationless (dry/gas-poor) merger alternatives of hierarchical clustering.

In the classical models of monolithic collapse (\citealt{eggen62}; \citealt{larson74a}; \citealt{larson75}; \citealt{carlberg84}; \citealt{arimoto87}), stars form in all regions during a rapid collapse and remain in their orbits, whereas the gas dissipates to the centre of the galaxy. The shape and ellipticity of stellar orbits are expected to depend on the initial angular momentum of the collapsing protogalactic gas cloud and on the effect of gas viscosity in redistributing angular momentum, towards outer radii, during the collapse. The sinking gas is continuously enriched by evolving stars. As a consequence, stars formed in the centre are predicted to be more metal-rich than those born in the outer galaxy regions. The efficiency of this process is proportional to the depth of the galaxy potential well, so that a strong correlation between metallicity gradient and galaxy mass is expected. The evolving stars are also responsible for the $\alpha$-element enrichment of the interstellar medium (ISM). 
To reproduce the supersolar abundance of $\alpha$-elements with respect to the iron-peak elements, observed in early-type galaxies, the models require that the collapse and star forming process occurred on timescales $\la$~1~Gyr, which will produce null or very small age gradients (\citealt{arimoto87}; \citealt{matteucci94}; \citealt{thomas99}). Recently, feedback processes such as supernova-driven galactic winds have been re-considered in more detailed numerical models of dissipative collapse (\citealt{larson74b}; \citealt{arimoto87}; \citealt{gibson97}; \citealt{martinelli98}; \citealt{chiosi02}; \citealt{kawata03}; \citealt{pipino06}; \citealt{pipino07}). It is suggested that they may create and shape the abundance gradients. Galactic winds initiate when the energy injected by supernovae explosion balances the gravitational binding energy of the ISM. The winds evacuate the gas, eliminating the essential fuel for any future star formation. The shallower local potential well of the external galactic regions supports the early development of winds with respect to the central regions. Therefore, in the central regions the star formation and the chemical enrichment last longer. Dissipation of gas towards the galaxy centre and a time-delay in the occurrence of galactic winds cooperate in steepening any metallicity gradient. In summary, dissipative collapse models predict: i) very steep negative metallicity gradients, that strongly correlate with galaxy mass (\citealt{chiosi02}; \citealt{kawata03}), ii) small or null age gradients, as a consequence of the short timescales involved in the collapse and star formation processes, iii) $\alpha$-element enhancement gradients that can be either positive, negative or null (\citealt{martinelli98}; \citealt{pipino06}), due to the ``\textit{interplay between local differences in the star formation timescale and gas flows}" (\citealt{pipino07}), iv) isophotal shape and ellipticity proportional to the protogalactic initial angular momentum and its radial redistribution, v) no peculiar internal kinematic structure. The difference between the original and revised dissipative collapse models focuses on the assembly of the initial gaseous material. The revised view accounts for the possibility of either a unique primordial cloud or the coalescence of many gaseous clumps without any preexisting stars. Otherwise, the physical processes in galaxy formation are similar to that described above.

In the hierarchical clustering scenario of galaxy formation (eg. \citealt{kauffmann93}; \citealt{cole94}; \citealt{baugh96}; \citealt{kauffmann98}; \citealt{delucia06}), elliptical galaxies are produced by merging events. Early numerical simulations of galaxy mergers predicted contradictory radial variations of stellar properties in merger remnants. \cite{white80} suggested that dissipationless mergers cause a flattening of metallicity gradients, whereas \cite{albada82} argued that the position of the stars in the local potential are preserved by violent relaxation and the gradients in the progenitors are only affected moderately. Further simulations of \cite{barnes91}, considered the hydrodynamical physics of the gas and found that during a dissipative merger a significant fraction of the progenitors' gas tends to migrate toward the central regions of the merger remnant. \cite{mihos94} showed that the gas accumulated in the central regions may trigger a local secondary burst of star formation. Galaxy merger-induced star formation is predicted to produce metallicity and age gradients. Positive or negative $\alpha$-element enhancement gradients may also originate, depending on the original enhancement of the gas and the duration of the burst (\citealt{thomasgre99}; \citealt{thokauf99}). \cite{bekki99} simulated dissipative mergers of two gas-rich disc galaxies of varying mass to form an elliptical galaxy remnant. Their models show that metallicity gradients in merger remnants are shallower with respect to those predicted by dissipative collapse models and only weakly dependent by the remnant galaxy mass. \cite{kobayashi04} simulated the formation and chemodynamical evolution of elliptical galaxies in a cold dark matter cosmology, focusing on internal metallicity gradients. The models ranged from a monolithic collapse scenario, seen as the assembly of tens of gas-rich subunits at high redshift, to dissipative and dissipationless mergers of equal-mass galaxies at low redshift (which she denoted ``major mergers''). The result was that galaxies of a given mass had steep metallicity gradients if formed by collapse and shallower gradients if formed by major merger (dissipative or dissipationless). 

Recently, combining the Millennium cosmological N-body simulation (\citealt{springel05}) with semi-analytic models, \cite{delucia06} and \cite{delucia07} focused on the distinction between the formation and assembly time (redshift) of elliptical galaxies and brightest cluster galaxies (BCGs) in the hierarchical clustering scenario. The models make use of the prescription of feedback from a central active galactic nucleus (AGN) and supernova (see \citealt{croton06}) to prevent excessive galaxy-mass growth from cooling flows. Their model predicts a mass-dependent evolutionary history (``down sizing'' scenario), with more massive galaxies (e.g BCGs, BGGs; brightest group galaxies) forming a large fraction of their stars (50~$-$~80 per cent) in progenitor systems at redshift $\sim$~2.5. The progenitors then assemble into a single final object at z~$\sim$~0.8. The formation epoch of less massive galaxies is predicted to be z~$\sim$~1.9 and the assembly time at redshift $\sim$~1.5. The peak of the star formation history of massive galaxies is at redshift $\sim$~5, and it progressively moves towards lower redshifts for less massive systems. Similarly, the duration of the star forming episode is predicted to last longer in low mass galaxies. Mergers are thought to induce bursts of star formation. Therefore, the star formation history of a galaxy presents the ``bursty'' behaviour described in \cite{delucia06}. 

The kinematics and internal structure of a merger remnant are a consequence of the progenitors' mass ratios and the efficiency of the dissipative process (e.g. \citealt{naab06}). \cite{naab03} show that a dissipationless binary merger of equal-mass disk galaxies leads to a slowly rotating and anisotropic supported system ($(v/\sigma)^{*}<$~0.4) with preferentially boxy isophotes and significant minor-axis rotation. Whereas, the merger of unequal-mass disk galaxies produces a rotationally supported elliptical galaxy with a small amount of minor-axis rotation ($(v/\sigma)^{*}=$~1.2; suggesting isotropic oblate rotators) and disky isophotes. In dissipative mergers the gas is found to settle at the galaxy centre deepening the potential well and making it more axisymmetric. The isophotal shape of an equal-mass merger is then strongly affected by the gas, since most of the stars are gravitationally induced to move from boxy to more elliptical or even disky orbits. Similarly, the disky isophotal shape of an unequal-mass merger remnant is weakly affected by the presence of gas (\citealt{naab06}). Furthermore, in simulations of an equal-mass merger with induced star formation (\citealt{bekki97}) gradual gas dissipation forms an elliptical galaxy with disky isophotes, whereas a merger with rapid star formation produces both boxy and disky isophotes depending on the viewing angle of the observer. The merger of unequal-mass gas-rich disk galaxies is found to exhaust a large amount of the gas owing to moderate star formation and ending in the formation of an S0 galaxy (\citealt{bekki98}). In summary, hierarchical galaxy formation models predict: i) shallow negative metallicity gradients, with little host mass dependence, ii) age gradients and iii) $\alpha$-element enhancement gradients, that can be either positive or negative, due to the duration and location of the merger-induced star formation and the original abundance pattern of the gas, iv) isophotal shape  and ellipticity depending on the progenitors mass ratio and the efficiency of gas dissipation, v) peculiar internal kinematic structures.

The contrasting predictions of the galaxy formation models underline the lack of complete understanding of the physical processes behind the formation of galaxies. Observationally, we now have the ability to obtain spatially resolved high signal-to-noise integrated spectra out to large galactic radii (and hence larger mass fractions). This allows us to consider the physical mechanisms acting locally and how their properties vary with radius (i.e. investigating beyond the central regions), proving invaluable constraints on the processes of galaxy formation and evolution. Recently, an increasing number of works have focused their attention on stellar population radial profiles of galaxies in different environments (e.g. \citealt{kobaari99}; \citealt{mehlert03}; \citealt{sanchez06}; \citealt{sanchez07}; \citealt{brough07}; \citealt{reda07}). The results underline a common behaviour for the stellar population radial profiles of the studied galaxies: early-type galaxies are often characterised by strong metallicity gradients, shallow age gradients and null or statistically insignificant $\alpha$-element enhancement gradients. 

Following from \cite{spola08a} (hereafter Paper~I) we probe a possible star formation and chemical evolutionary history of the early-type galaxies NGC~1407 and NGC~1400. We accomplish the task using the spectral Lick/IDS indices (\citealt{faber85}; \citealt{worthey94}) to determine the radial profiles in age, metallicity, [Z/H], and $\alpha$-element abundance ratios, [$\alpha$/Fe], out to $\sim$~0.6 (NGC~1407) and $\sim$~1.30 (NGC 1400) times the galaxies' effective radii ($r_{e}$; the radius within which half the galaxy light is contained). Central values and gradients in age, [Z/H] and [$\alpha$/Fe] are estimated and used as proxies of the quantity, velocity and duration of gas dissipation, star formation and possible merger events during galaxy formation. Furthermore, we have considered the results from the spatially resolved radial kinematics and surface photometry analysis performed in Paper~I.
We compared colour radial profiles with colours predicted from spectroscopically derived ages and metallicities using single stellar population models.

This paper is organised as follows. In Section~2 we describe the two sample galaxies. In Sections~3 and 4 we describe the observations and the relevant data reduction. Section~4 presents the spatially resolved stellar population parameters. In Section~5 observed and predicted colour index radial profiles are examined. In Section~6 we summarise our results, and discuss the possible star formation and evolutionary histories of NGC~1407 and NGC~1400. In Section~7 conclusions are presented.

\section{THE DATA SAMPLE}
The NGC~1407, or Eridanus-A, group is one of the 60
groups studied as part of the Group Evolution Multi-wavelength Study (GEMS; \citealt{osmond04}; \citealt{forbes06}). The study was designed to probe the evolution of groups and their member galaxies. NGC~1407 is the brightest group galaxy of the NGC~1407 (or Eridanus-A) group. NGC~1400 is the second brightest group galaxy and it is characterised by a peculiar low velocity. A complete description of the data sample is presented in Paper~I. Table~\ref{ga_prop} reports some of the galaxies' properties.

\begin{table}
\begin{center}
\begin{tabular}{ccccccccccc}
\hline 
\hline 
Name & Type & PA & $r_{e}$ & M$_{B}$ & M$_{K}$ \\ 
     &      &(degrees) &(arcsec) &(mag) &(mag) \\ 
\hline 
NGC~1407 & E0 &35 &72$\arcsec$ & $-$21.22 & $-$24.77 \\ 
NGC~1400 & S0 & 40 &27$\arcsec$ & $-$19.95 & $-$23.74 \\ 
\hline \hline
\end{tabular}
\end{center}
\caption{Galaxies properties. Type: morphological type extracted from NASA/IPAC Extragalactic Database (NED); PA: position angle of the major axis; $r_{e}$: effective radius from HyperLEDA; M$_{B}$, M$_{K}$: absolute B-band and K-band magnitudes obtained from HyperLEDA and 2MASS, respectively.}
\label{ga_prop}
\end{table}


\section{OBSERVATIONS AND DATA REDUCTION}
Spectral observations were performed with the ESO Faint Object Spectrograph
and Camera (EFOSC2) mounted on the ESO 3.6m telescope at La Silla
Observatory, Chile. Table~\ref{obs_conf} summarises the instrumental configuration adopted during the observing run. Lick/IDS and spectrophotometric standard stars were taken at the
parallactic angle, the latter with a 5$\arcsec$ wide slit. The Lick/IDS stars are also used
as velocity standards.

\begin{figure}
\begin{center}
\includegraphics[scale=.43]{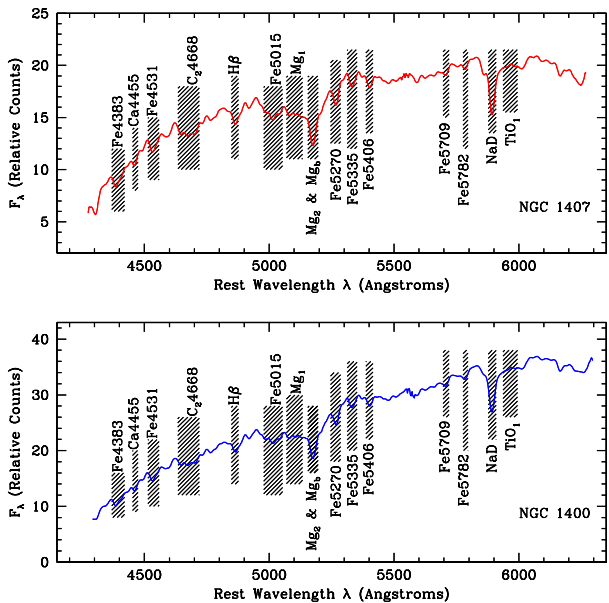}
\caption[spectra]{The spectrum of NGC~1407 and NGC~1400, obtained from co-adding the individual 1D spectra. The flux scale is arbitrary. The shadowed regions indicate the central band of the Lick/IDS indices used in the stellar populations analysis.}
\label{spectra}
\end{center}
\end{figure}

Data reductions were carried out with IRAF\footnote{IRAF is
distributed by the National Optical Astronomy Observatories, which are
operated by the Association of Universities for Research in Astronomy,
Inc., under cooperative agreement with the National Science
Foundation.} using the procedures adopted in Paper~I. In Fig.~\ref{spectra} is presented the spectrum of NGC~1407 and NGC~1400, obtained from co-adding the individual 1D spectra.

We spatially resolved 68 and 82 apertures along the observing axes for
NGC~1407 and NGC~1400, respectively. The spatial width (i.e. the number of CCD rows binned) for each
extracted aperture increases with radius to achieve a signal-to-noise
ratio of $\sim$~30~{\AA}$^{-1}$ at 5000~{\AA}. We reached a radial extent of $\sim$~0.56$r_{e}$ ($\sim$~4.11 kpc) for NGC~1407 and $\sim$~1.30$r_{e}$ ($\sim$~3.58 kpc) for NGC~1400.

\begin{table}
\begin{center}
\begin{tabular}{lc}
\hline 
\hline 
Telescope                   & ESO-3.6m         \\
Spectrograph                & EFOSC2           \\
Chip size (pixels)          & 2048 x 2048      \\
Grism resolution  FWHM      & 7.8 {\AA}         \\ 
Pixel scale                 & 0.314$\arcsec$  \\
Binning                     & 2 x 2            \\
Slit  width                 & 1.2$\arcsec$          \\ 
Wavelength coverage         & 4320 - 6360 {\AA} \\ 
Seeing FWHM                 & $\sim$ 1$\arcsec$    \\
Exposure time               & 3 x 1200 s        \\
NGC~1407 Slit P.A.          & 44$^{\circ}$     \\
NGC~1400 Slit P.A.          & 42$^{\circ}$     \\
\hline
\hline
\end{tabular}
\end{center}
\caption{Spectral observing parameters.}
\label{obs_conf}
\end{table}

\section{SPATIALLY RESOLVED STELLAR POPULATIONS}
The wavelength range of 4300-6300{\AA} covered by our observations includes 16 Lick/IDS indices (\citealt{worthey94}). These include one Balmer index (H$\beta$), three Magnesium indices (Mg$_{1}$, Mg$_{2}$ and Mg$_{b}$), Calcium (Ca4455), Sodium (NaD), Titanium Oxide (TiO$_{1}$) indices plus nine Iron indices Fe4383, Fe4531, Fe4668 (referred as C4668), Fe5015, Fe5270, Fe5335, Fe5406, Fe5709 and Fe5782. The indices were measured using the method described in \cite{proctor02}. \cite{trager98} index definitions were adopted in the analysis. The calibration procedure of our indices measurements to the Lick/IDS system is described in Appendix~\ref{lickcalib}.

Fully corrected Lick/IDS line-strength indices, extracted from the aperture spectra along the observing axis of both the galaxies, are available electronically. In Figures~\ref{n1407multirad} and \ref{n1400multirad} the Lick/IDS index values are plotted against the galactocentric radius of NGC~1407 and NGC~1400 respectively. Logarithmic gradients of the indices were measured by applying the linear regression analysis described in Appendix~\ref{lsf}. The results are reported in Tables~\ref{line1} (NGC~1407) and \ref{line2} (NGC~1400).

\subsection{Stellar population model fitting procedure}
\label{stepop}
The observed  Lick/IDS indices have been used to derive luminosity-weighted log(Age), total metallicity [Z/H] and $\alpha$-abundance ratio [$\alpha$/Fe]. The $\alpha$-abundance ratio is parametrised by [E/Fe]; the latter quantifies the enhancement from the $\alpha$-elements (N, O, Mg, Na, Si, Ti) with respect to the Fe-peak elements (Cr, Mn, Fe, Co, Ni, Cu, Zn; see \citealt{thomas03} for details). 

Briefly, the technique (\citealt{proctor02}) involves  the  simultaneous  comparison  of  as  many  observed indices  as possible to single stellar  population (SSP) models. The best fit is achieved throughout a statistical $\chi^{2}$-minimisation technique, which reduce the deviations $\chi^{2}$ between  the observed  and the  modelled values  as a  fraction  of the index errors, i.e. $\chi^{2}$. The strength of the method is that it works with as many indices as possible in order to break the age-metallicity degeneracy that affects each index differently .

In this work, we use \cite{thomas03} SSP models. Models are provided with: [$\alpha$/Fe] = $-$0.3, 0.0, 0.3, 0.5; ages from 0.1 to 15 Gyr ($-$1 $\leq$ log(age) $\leq$ 1.175, in steps of 0.025 dex); [Z/H] from $-$2.25 to 0.8 in steps of 0.025 dex.

\begin{figure}
\begin{center}
\includegraphics[scale=0.8]{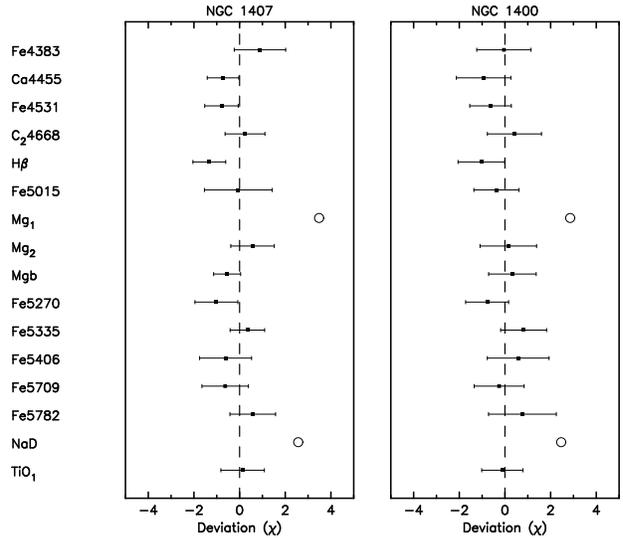}
\caption{The average deviation in units of error (i.e. $\chi$) for the 16 observed Lick/IDS
indices. Indices are listed in increasing wavelength order. Error bars represent the rms scatter in the deviations. Deviations after the removal of the poorly fitted Mg$_{1}$ and NaD indices (open circles) and of the indices omitted by the iterated 3$\sigma$-clipping routine (see text for details).}
\end{center}
\label{multichi2}
\end{figure}

We began the model fitting process obtaining the best fit to all 16 observed indices. The residuals are expressed in terms of index errors $\chi$. We found that Mg$_{1}$ and NaD indices are poorly fit by the \cite{thomas03} SSP models. Indeed, it is well known that interstellar absorption affects the strength of the NaD index. The Mg$_{1}$ index shows a small rms deviation but has an average deviation value greater then 3$\sigma$. This deviation is not caused by the 5102{\AA} sky emission line, however a large average deviation could be associated with poor flux calibration. In fact, the Mg$_{1}$ and Mg$_{2}$ indices are known to be sensitive to flux calibration corrections. We detected some residuals from the 5577{\AA} sky emission line, but they did not affect any nearby Lick/IDS absorption features. We performed a new fit removing the Mg$_{1}$ and NaD indices and applying an iterated 3$\sigma$-clipping function (\citealt{proctor04b}; \citealt{proctor05}) for the remaining indices. This lead to a better fit with improved residuals (Fig.~2).

\subsection{Central stellar population parameters}
The central stellar population parameters of NGC~1407 and NGC~1400 were measured by averaging the values of all the apertures within a radius of $r_{e}/8$. Central values of age, total metallicity, [E/Fe] ratio and velocity dispersion are reported in Table~\ref{centralval} together with values from previous studies.

We find that our measurements of NGC~1407 are in good agreement with literature values with the exception of the results of \cite{denicolo05} and \cite{thomas05}. The former found a younger central age (2.5 Gyr), a more metal rich stellar population ([Z/H]=0.67) and less enhancement from $\alpha$-elements ([E/Fe]=0.18). The reason for this disagreement may be associated with the emission line corrections they applied. In \cite{thomas05} the central age is younger (7.4 Gyr) than ours, whereas the [Z/H]=0.38 and [E/Fe]=0.30 values are in good agreement. For NGC~1400 our values are in good agreement with \cite{howell05} and \cite{thomas05}, but only partially with \cite{barr07}. In the latter, the central age is younger, the [E/Fe] ratio is in good agreement, but the total metallicity value is higher than our value. Thus, we find good agreement with the majority of literature values. 

In summary, we find that the central regions of both the galaxies present old ($\geq$~10~Gyr), metal-rich stellar populations with supersolar abundances of $\alpha$-elements.

\begin{table*}
\begin{center}
\begin{tabular}{lcccc}
\hline   
\hline
                   & Age$_{0}$           &  [Z/H]$_{0}$             &  [E/Fe]$_{0}$       &   $\sigma_{0}$  \\
                   & (Gyr)               &  (dex)                   &  (dex)            & (km s$^{-1}$)     \\
\hline
NGC~1407 & && \\
\hline
This work          & 12.0 $\pm$ 1.1       & 0.29 $\pm$ 0.08         & 0.41 $\pm$ 0.05  &    278.9 $\pm$ 4.3       \\
\cite{humbuote06}  & 12.0 $\pm$ 2.0       & 0.35$^{\dag}$ $\pm$ 0.06         & 0.33 $\pm$ 0.02  &    259.7 $\pm$ 3.7        \\
\cite{cenarro07}   & 11.9 $\pm$ 1.4       & 0.38 $\pm$ 0.04         & 0.33 $\pm$ 0.01  &    293.5 $\pm$ 4.1        \\
\cite{cantiello05} & 11.0 $\pm$ 1.0       &      $-$                &      $-$         &           $-$             \\
\cite{howell05}    & 9.5  $\pm$ 2.2       & 0.56 $\pm$ 0.07         & 0.30 $\pm$ 0.04  &    296.0 $\pm$ 4.0        \\
\cite{zhang07}     & 9.2  $\pm$ 0.7       &      $-$                &      $-$         &           $-$             \\
\cite{thomas05}    & 7.4  $\pm$ 1.8       & 0.38 $\pm$ 0.04         & 0.30  $\pm$ 0.02 &    259.7 $\pm$ 3.7        \\
\cite{denicolo05}  & 2.5  $\pm$ 0.6       & 0.67 $\pm$ 0.05         & 0.18  $\pm$ 0.05 &    265.0 $\pm$ 17.0       \\
\hline
NGC~1400& & & \\
\hline
This work          & 13.8  $\pm$ 1.1      & 0.25 $\pm$ 0.06         & 0.33 $\pm$ 0.04  & 254.4 $\pm$ 3.1       \\
\cite{howell05}    & 14.2  $\pm$ 3.0      & 0.31 $\pm$ 0.13         & 0.35 $\pm$ 0.04  & 285.0 $\pm$ 5.0         \\
\cite{thomas05}    & 15.0  $\pm$ 1.6      & 0.23 $\pm$ 0.03         & 0.34 $\pm$ 0.01  & 256.2 $\pm$ 3.0         \\
\cite{barr07}      & 9.57  $\pm$ 1.1      & 0.66$^{\dag}$ $\pm$ 0.03         & 0.34 $\pm$ 0.01  & 215 $\pm$ 7.0           \\
\hline 
\hline
\end{tabular}
\end{center}
\caption{Central (r $<$ r$_{e}/8$) stellar population parameters for NGC~1407 and NGC~1400 from this work and from previous studies. The errors are on the mean. The symbol $\dag$ indicates [Z/H] values that were not provided in the original works, but that we recovered using the [Z/H] = [Fe/H] + 0.94[$\alpha$/Fe] (\citealt{trager00a}).}
\label{centralval}
\end{table*}

\subsection{Stellar population radial profiles}
\begin{figure*}
\includegraphics[scale=.80]{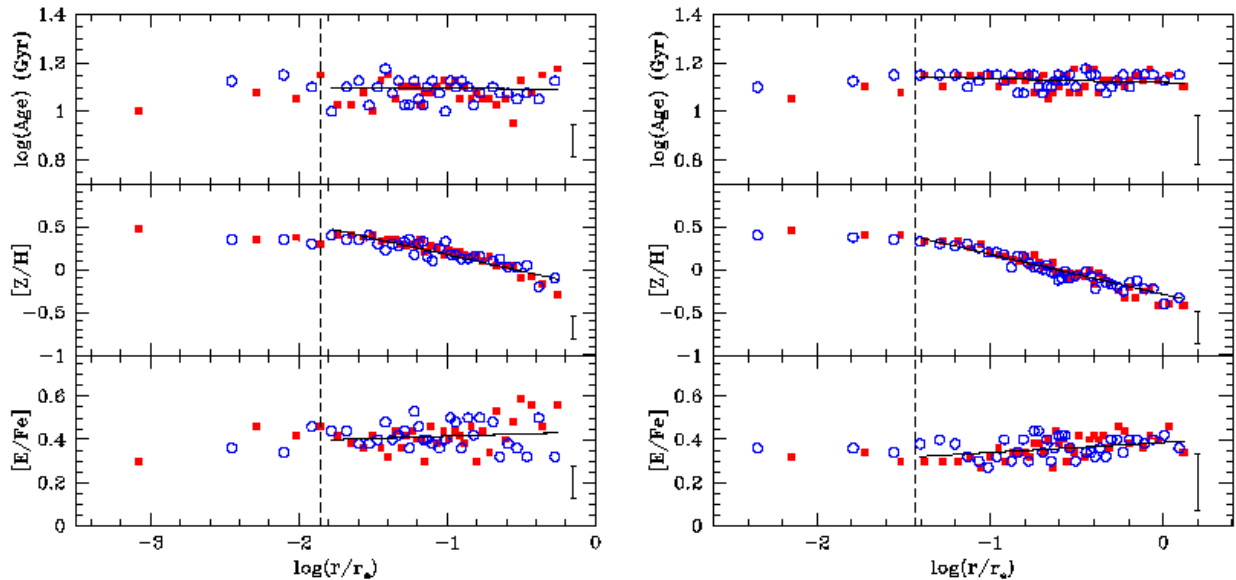}
\caption{Spatially resolved radial profiles for the NGC~1407 (left panel) and NGC~1400 (right panel) derived stellar population parameters: log(Age), total metallicity [Z/H] and $\alpha$-elements abundance ratio [E/Fe]. Filled red squares represent values from the spatially resolved apertures extracted along the observing axis in the negative binning direction; open circles are the same in the positive binning direction (see text for details). The approximate extent of the seeing disc is identified by a dashed line. The error bars express the median error in the derived parameters. Gradients (black thick lines) are measured excluding the points inside the region affected by the seeing. The quantities are measured in logarithmic radius scaled by the effective radius of the galaxy.}
\label{pop}
\end{figure*}

Stellar population parameters extracted from the spatially resolved apertures along the observing axes of NGC~1407 and NGC~1400 have been plotted in Fig.~\ref{pop}. The logarithmic gradients of the radial profiles are measured using the same method adopted for Lick/IDS line-strength radial profiles (see Appendix~\ref{lsf}), beyond the seeing limit of 1 arcsec; the results are summarised in Table~\ref{gradients}. The quoted errors are the standard errors on the slope.

The radial age profile of NGC~1407 indicates a constant old age for the spatially resolved regions of the galaxy. The measured age gradient is not statistically significant: $-$0.00~$\pm$~0.02 dex~per~radius~dex. The region inside 5~arcsec was analysed to search for variations in the age profile likely to be associated with the possible kinematically decoupled core (Paper~I). No significant substructure was found that could not be due to the intrinsic scatter of the data points. The measured age gradient of NGC~1400 is also not statistically significant: $-$0.02~$\pm$~0.02~dex~per~radius~dex. Previous works on age gradients of early-type galaxies in different environments (see Table~\ref{gradients}) are consistent with zero-slope or very shallow gradients. \cite{mehlert03} found no significant age gradients for 91 per cent of their sample of 35 early-type galaxies in the Coma cluster. \cite{sanchez06}, in a study of 76 early-type galaxies situated in low- and high-density environments did not find any overall significant age gradient in the sample. \cite{reda07} also found zero-slope age gradients for 12 isolated early-type galaxies. In a recent study, \cite{sanchez07} found a shallow mean age gradient for 11 early-type galaxies covering a wide range in luminosity and environment.

Both NGC~1407 and NGC~1400 galaxies present metallicity radial profiles similar in shape and slope. We measured gradients of $-$0.38~$\pm$~0.04 and $-$0.47~$\pm$~0.04~dex~per~radius~dex, respectively. \cite{kobaari99} derived a mean metallicity gradient from absorption-line strengths literature values of 80 early-type galaxies of $-$0.30~$\pm$~0.15~dex~per~radius~dex. \cite{sanchez06} found a metallicity gradient of $-$0.21~$\pm$~0.02~dex~per~radius~dex for low-density environment elliptical galaxies and $-$0.33~$\pm$~0.06~dex~per~radius~dex for high-density environment elliptical galaxies. Similar negative metallicity values were found in \cite{sanchez07} ($-$0.33~$\pm$~0.07~dex~per~radius~dex) and \cite{reda07} ($-$0.25~$\pm$~0.05~dex~per~radius~dex) for a sample of isolated galaxies. Our literature search (Table~\ref{gradients}) finds a range of variation for the measured metallicity gradients: i.e going from a minimum of grad~[Z/H]~=~$-$0.16~dex~per~radius~dex (in the Coma cluster galaxies; \citealt{mehlert03}) to a maximum of $-$0.34~dex~per~radius~dex (for BGGs and BCGs; \citealt{brough07}).

Scatter along the fit is seen in both the [E/Fe] profiles, but no statistically significant gradient was measured. Statistically no significant radial gradients of [E/Fe] were found by \cite{mehlert03} with an average of 0.05~$\pm$~0.05~dex~per~radius~dex. Null mean gradients were found by \cite{sanchez07}, \cite{reda07} and \cite{brough07}.  Given the absence of significant gradients in the $\alpha$-abundance ratio values, those derived from the central regions (r $\leq$ $r_{e}/8$; Table~\ref{centralval}) are representative of the whole galaxy. Of particular interest is that both galaxies are characterised by a significant supersolar abundance of $\alpha$-elements, [E/Fe]$_{0}$~=~+0.41 and +0.33 respectively, suggesting a short star formation time scale. The [E/Fe] radial profiles, as well as the metallicity radial profiles and age, did not present any evidence substructure related to the possible 
kinematically decoupled core in NGC~1407 (Paper~I).

In summary, we find that the stellar population radial profiles of both NGC~1407 and NGC~1400 galaxy are similar in shape and slope, suggesting a similar formation and evolution scenario. The radial age and [E/Fe] profiles are flat, with no statistically significant gradients measured. The metallicity radial profiles present significant negative gradients, with an average value steeper than previous literature works on metallicity gradients of early-type galaxies.

\begin{table*}
\begin{center}
\begin{tabular}{cccccc}
\hline   
\hline
               &N & Type \& Environment & grad Age           &  grad [Z/H]       &  grad [E/Fe] \\
               &  &                     & (dex per dex)   &(dex per dex)   &(dex per dex) \\
\hline
NGC~1407       & 1 & E; group                     &$-$0.00 $\pm$ 0.02    &$-$0.38 $\pm$ 0.04   &0.02 $\pm$ 0.02\\
NGC~1400       & 1 & S0; group                    &$-$0.02 $\pm$ 0.02    &$-$0.47 $\pm$ 0.04   &0.04 $\pm$ 0.03\\
\hline 
\textbf{Averages}& &               &                    &                     &                 \\
This work      & 2 & E, S0; group  & $-$0.01 $\pm$ 0.02   &  $-$0.42 $\pm$ 0.04   &  0.03 $\pm$ 0.02 \\

\cite{gorgas90} &15 &Es, S0s, BCGs; groups, clusters &      $-$           &  $-$0.22 $\pm$ 0.10   &    $-$          \\
\cite{fisher95} &16 &BCGs, Es; groups, clusters  &      $-$           &  $-$0.25 $\pm$ 0.10   &    $-$          \\
\cite{kobaari99}&80 &Es; groups, clusters &      $-$           &  $-$0.30 $\pm$ 0.15   &    $-$          \\
\cite{mehlert03}& 35 &Es, S0s; Coma cluster &  0.43 $\pm$ 0.46   &  $-$0.16 $\pm$ 0.12   & 0.05 $\pm$ 0.05 \\
\cite{sanchez06}& 61 &Es, S0s; field, group, Virgo cluster  &  0.08 $\pm$ 0.02 &  $-$0.21 $\pm$ 0.02 &  $-$    \\
\cite{sanchez06}& 15 &Es, S0s; Coma cluster, Abell clusters  &  0.03 $\pm$ 0.07 &  $-$0.33 $\pm$ 0.06 &  $-$    \\
\cite{sanchez07}& 11 &Es, S0s; field, groups, Leo group, Virgo cluster&   0.16 $\pm$ 0.05  &  $-$0.33 $\pm$ 0.07   & 0.00 $\pm$ 0.05 \\
\cite{brough07} & 6  &BGGs, BCGs; groups, clusters &   0.01 $\pm$ 0.04  &  $-$0.34 $\pm$ 0.08   & 0.06 $\pm$ 0.02 \\
\cite{reda07}   & 12 &Es, S0s; isolated &   0.04 $\pm$ 0.08  &  $-$0.25 $\pm$ 0.05   & $-$0.03 $\pm$ 0.02 \\
\hline
\hline
\end{tabular}
\end{center}
\caption{Stellar population radial gradients for NGC~1407 and NGC~1400. Average values of age, [Z/H] and [E/Fe] gradients in early-type galaxies obtained in other works. N: number of galaxies considered in the study. Type \& Environment: morphological type and environment of the studied galaxies. Grad~Age: age gradient, $\frac{\Delta log(Age)}{\Delta log(r)}$. Grad~[Z/H]: total metallicity gradient,  $\frac{\Delta [Z/H]}{\Delta log(r)}$. Grad~[E/Fe]: $\alpha$-elements abundance ratio gradient, $\frac{\Delta [E/Fe]}{\Delta log(r)}$. The quoted errors are the standard errors on the slope.}
\label{gradients}
\end{table*}

\section{Observed and predicted colour index radial profiles}
\label{colorcomp}
In Paper~I we presented the results of a spatially resolved surface photometry analysis of NGC~1407 and NGC~1400. The study was performed using HST/ACS and Subaru Suprime-Cam multi-band imaging data. The high instrumental resolution of HST/ACS allowed us to sample NGC~1407 over radii $\sim$~0.1~$-$~100~arcsec ($\sim$~0.01~$-$~10.20~kpc) in B and I band. The wide-field of Subaru Suprime-Cam allowed us to sample NGC~1407 and NGC~1400 over radii $\sim$~5~$-$~100~arcsec ($\sim$~0.51~$-$~10.20~kpc) in g', r' and i' Sloan band. The lost of the inner $\sim$~5~arcsec is caused by saturation within the CCD (see Paper~I for details).

\begin{figure*} \includegraphics[scale=0.4, angle=0]{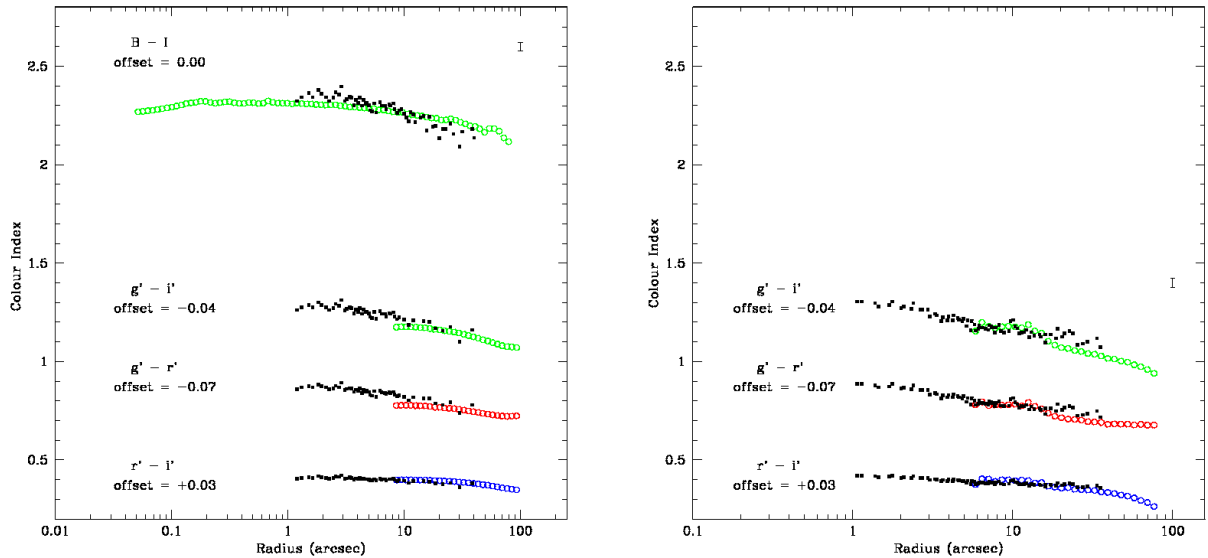}
\caption[]{Observed (open circles) and predicted (filled black squares) colour profiles for NGC~1407 (left panel) and NGC~1400 (right panel). Colours are predicted from the measured ages and metallicities using the \cite{TMK04} SSP models. The error bar is the median value for the predicted parameters. Offsets have been applied to the predicted data to match our results.}  
\label{col}
\end{figure*}

In Fig.~\ref{col} we super-imposed the colour predictions from the SSP models of \cite{thomas03} (http://www-astro.physics.ox.ac.uk/$\sim$maraston/SSPn/colors/) to the observed colour profiles. The models provides colours as a function of age and metallicity. Therefore, using our spectroscopically measured ages and metallicities (see Section~4.3) we created radial profiles of predicted colours.

The predicted g'$-$i', g'$-$r' and r'$-$i' profiles of both the galaxies required \textit{ad hoc} vertical offsets to match the observed colours: i.e. $-$0.04, $-$0.07 and +0.03, respectively. We compared observed and predicted colours by measuring the radial gradient of the profiles. The results are reported in Table~\ref{colgrad1}. The predicted NGC~1407 B$-$I profile presents a steep gradient that is not seen in the observed colours; whereas small differences are found between the Sloan colour profile gradients. In a similar analysis on NGC~821, \cite{proctor05} also found it necessary to apply offsets in order to match observations. These offsets might be explained with effects of internal extinction, although systematic effects (e.g. SSP modelling errors) may be present. 

To better investigate the difference in slope we calculated the ratio $\Delta C / \Delta [Z/H]$ (see Table~\ref{colgrad2}). In particular, $(\Delta C / \Delta [Z/H])_{Obs}$ is the ratio between the observed colour index gradient and the gradient in metallicity that we found from the Lick/IDS indices analysis (see Table~\ref{gradients}). In the $(\Delta C / \Delta [Z/H])_{SSP}$ ratio, $\Delta C$ expresses the colour index difference estimated from the SSP model assuming a constant age (12 Gyr for NGC~1407 and 13.8 Gyr for NGC~1400; see Table~\ref{centralval}) and metallicity values of [Z/H]~=~0, [Z/H]~=~+0.38 for NGC~1407 and [Z/H]~=~0, [Z/H]~=~+0.47 for NGC~1400; consequently, $\Delta [Z/H]$ is the difference between the two metallicity values.
The assumption of constant age is supported by the fact that we do not observe any age gradient in the stellar population (see Fig.~\ref{pop}). We found that the discrepancy between observed and predicted g'$-$i', g'$-$r', r'$-$i' values is small but it becomes statistically significant for the B~$-$~I colour index.

In summary, both galaxies demonstrate the observed trend typical of luminous elliptical galaxies, with the galaxy centre redder than the outer regions (e.g. \citealt{franx89}). The SSP predictions for the Sloan colours match the observations if \textit{ad hoc} offsets are applied. On the other hand, we found a discrepancy in the slope between the NGC~1407 observed and predicted radial profile for the B$-$I colour.

\begin{table}
\begin{center}
\begin{tabular}{ccccc}
\hline
\hline
NGC 1407   & Profile range & Obs grad & SSP grad\\
           &  (arcsec)     & (mag dex$^{-1}$) & (mag dex$^{-1}$) \\
\hline
B~$-$~I    & 0.1 $-$ 100 & $-$0.16 $\pm$ 0.01 & $-$\\
           & 1 $-$ 40    & $-$0.05 $\pm$ 0.01 & $-$0.16 $\pm$ 0.01\\
g'~$-$~i'  & 5 $-$ 100   & $-$0.12 $\pm$ 0.01 & $-$\\
           & 5 $-$ 40    & $-$0.10 $\pm$ 0.01 & $-$0.14 $\pm$ 0.01\\
g'~$-$~r'  & 5 $-$ 100   & $-$0.06 $\pm$ 0.01 & $-$\\
           & 5 $-$ 40    & $-$0.08 $\pm$ 0.01 & $-$0.10 $\pm$ 0.01\\
r'~$-$~i'  & 5 $-$ 100   & $-$0.06 $\pm$ 0.01 & $-$\\
           & 5 $-$ 40    & $-$0.04 $\pm$ 0.01 & $-$0.04 $\pm$ 0.01\\
\hline
\hline
NGC 1400   & Profile range & Obs grad & SSP grad\\
           &  (arcsec)     & (mag dex$^{-1}$) & (mag dex$^{-1}$) \\
\hline
g'~$-$~i'  & 5 $-$ 100   & $-$0.22 $\pm$ 0.01 & $-$\\
           & 5 $-$ 36    & $-$0.18 $\pm$ 0.01 & $-$0.15 $\pm$ 0.01\\
g'~$-$~r'  & 5 $-$ 100   & $-$0.11 $\pm$ 0.01 & $-$\\
           & 5 $-$ 36    & $-$0.12 $\pm$ 0.01 & $-$0.12 $\pm$ 0.01\\
r'~$-$~i'  & 5 $-$ 100   & $-$0.11 $\pm$ 0.01 & $-$\\
           & 5 $-$ 36    & $-$0.06 $\pm$ 0.01 & $-$0.04 $\pm$ 0.01\\
\hline
\hline
\end{tabular}
\end{center}
\caption{Observed and predicted colour radial gradients for NGC~1407 and NGC~1400.}
\label{colgrad1}
\end{table}

\begin{table}
\begin{center}
\begin{tabular}{ccc}
\hline
\hline
NGC 1407   & $(\Delta C /\Delta [Z/H])_{Obs}$ & $(\Delta C /\Delta [Z/H])_{SSP}$\\
\hline
B~$-$~I    & $-$0.13 $\pm$ 0.02  & $-$0.40 $\pm$ 0.01  \\
g'~$-$~i'  & $-$0.26 $\pm$ 0.02  & $-$0.35 $\pm$ 0.01  \\
g'~$-$~r'  & $-$0.21 $\pm$ 0.02  & $-$0.25 $\pm$ 0.01  \\
r'~$-$~i'  & $-$0.11 $\pm$ 0.02  & $-$0.10 $\pm$ 0.01  \\
\hline
\hline
NGC 1400   & $(\Delta C /\Delta [Z/H])_{Obs}$ & $(\Delta C /\Delta [Z/H])_{SSP}$\\
\hline
g'~$-$~i'  & $-$0.38 $\pm$ 0.02   & $-$0.36 $\pm$ 0.01   \\
g'~$-$~r'  & $-$0.26 $\pm$ 0.02   & $-$0.28 $\pm$ 0.01   \\
r'~$-$~i'  & $-$0.13 $\pm$ 0.02   & $-$0.09 $\pm$ 0.01   \\
\hline 
\hline
\end{tabular}
\end{center}
\caption{Observed and predicted $\frac{\Delta C}{\Delta [Z/H]}$ ratio. See Section~5 for details.}
\label{colgrad2}
\end{table}

\section{DISCUSSION}
To summarise, we have performed a detailed spectrophotometric study of the early-type galaxies NGC~1407 and NGC~1400. From spectral Lick/IDS line-strength indices we have measured radial ages, metallicities and abundance ratios. Results from high resolution wide-field multi-band imaging data (Paper~I) have been used to derive radial colour profiles.

The lack of radial variation in the luminosity-weighted mean age and $\alpha$-element profiles suggest that the central and external galaxy regions formed in a single burst of star formation on a timescale of $\sim$~1~Gyr and then evolved passively thereafter. The old central age values, 12.0~$\pm$~1.1 Gyr (NGC~1407) and 13.8~$\pm$1.1 Gyr (NGC~1400), constrain the redshift of the star forming episode to z~$\geq$~5. Thus neither NGC~1407 nor NGC~1400 have experienced any recent star formation, at least in a burst strong enough to significantly affect the observed integrated light. \cite{donas06} combined GALEX UV and optical photometry to study the colours of early-type galaxies. The sample included NGC~1407 and NGC~1400. UV colours are more sensitive to small bursts of star formation. They found the two galaxies to be old with no evidence of residual star formation activity. A similar result, also using GALEX, was found by \cite{yi05}. The uniformity of the age profiles implies that the galaxies have not experienced gaseous mergers since formation at redshift z~$\geq$~5. However, we can not preclude dissipationless merger events where starbursts are not triggered. 
A signature of peculiar dynamics associated with a merger event, might be represented by the possible kinematically decoupled core detected in NGC~1407 (Paper~I). However, the detection is uncertain and possibly due to a misalignment of the slit respect the nucleus of the galaxy. Moreover, the results of our galaxy isophote modelling show no evidence for recent mergers (see Paper~I for details).

The formation of the bulk of the stars at high redshift is in concordance with the predictions of the revised monolithic collapse models, whereas hierarchical models tend to place the formation of elliptical galaxy stars at lower redshift, e.g. z~$<$~2. Nevertheless, the recent numerical simulations of \cite{delucia06} pointed out the distinction between formation time of the stars in the progenitors, and the assembly time of the final elliptical galaxy, with particular emphasis on BCGs and BGGS (\citealt{delucia07}). Our results are in general agreement with their predictions for the peak of the star formation at redshift z~$\sim$~5 ($\sim$~12~Gyr) and the formation of $\sim$~70 per cent of the stars by z~$\sim$~2 ($\sim$~10~Gyr). However, we find that our results do not reproduce the predicted ``bursty'' behaviour for the star formation history, due to the assumption that bulge stars form during merger-induced bursts. We also find a discrepancy in the duration of the star formation episode (see below), predicted to last for several Gyr.

Recently, \cite{pipino07} has modelled the [$\alpha$/Fe] radial trend by an outside-in dissipative formation. It is a consequence of the interplay between the velocity of the $\alpha$-enhanced radial flows and the intensity and duration of the star formation process at any radius. Thus, the observed flatness of the NGC~1407 and NGC~1400 [$\alpha$/Fe] gradients should be considered a consequence of the faster inflow rate of the $\alpha$-enhanced gas with respect to the rate at which the gas is consumed by star formation, resulting in a pile-up of gas in the centre. \cite{thomas05} suggested that the [$\alpha$/Fe] ratio is proportional to the logarithm of the timescale of a star forming episode, which is the FWHM of a Gaussian star formation rate. The link was previously shown by \cite{thomasgre99} and it is in very good agreement with the chemodynamical simulations of \cite{pipino04}. They suggest that a ratio [$\alpha$/Fe]~=~0.2 for a composite stellar population requires formation timescales $\la$~1~Gyr. The values that we found suggest star formation episodes of a time-duration less than or equal to 1~Gyr (star formation rates $\geq$~100~M$_{\sun}$/year).

We next turn our attention to the strong metallicity gradients present in NGC~1407 and NGC~1400. The study of such gradients provide an additional diagnostic into the mechanism of galaxy formation and evolution, and a key constraint to discern between the different theoretical models. The measured [Z/H] gradient is $-$0.38~$\pm$~0.04 for NGC~1407 and $-$0.47~$\pm$~0.04 for NGC~1400. Such strong gradients are inconsistent
with the predictions of \cite{kobayashi04} for any of the dissipative or dissipationless merger models. In these models, small or null gradients are predicted due to the ``washing-out'' of the ambient gradients in the turbulent mixing of merging events. If star formation happens the central metallicity would increase slightly, creating a shallow gradient. The gradients in our galaxies are steeper than are achieved in either dissipative or dissipationless merger models. In addition, we can preclude that the galaxies have experienced a recent major merger due to the lack of tidal tails, shells and plumes expected to persist for $\sim$~3~Gyr after a major merger. At the same time, we can exclude the alternative of a minor merger: i.e. the accretion of a small, gaseous satellite galaxy (as defined by \citealt{kobayashi04}). In this scenario, the satellite galaxy is expected to contain low-metallicity gas and therefore high central metallicities are hard to achieve. On the other hand, steep metallicity gradients are a natural outcome of the revised dissipative collapse models. The steepness of the slope is a consequence of gas dissipation towards inner radii and the time-delay in the occurrence of galactic winds due to the difference in the galaxy local binding energy. 

\cite{cenarro07} performed a Lick/IDS spectral analysis of a small sample of globular clusters in NGC~1407. We calculated the mean value of the derived stellar populations for the red (metal-rich) globular cluster subpopulation, which have been shown to be most closely associated with the underlying galaxy light (e.g. \citealt{forbes04}; \citealt{forbes06c}; \citealt{norris07}). We find: mean age~=~10.2~$\pm$~2.1 Gyr; [Z/H]~=~$-$0.17~$\pm$~0.08; [$\alpha$/Fe]~=~+0.27~$\pm$~0.07. In order to compare these values at a similar galactocentric radius, we slightly extrapolate our stellar population profiles (Fig.~\ref{pop}) to 1 effective radius (the average value for the globular cluster locations) finding: mean age~=~12.5~$\pm$~1.1 Gyr; [Z/H]~=~$-$0.21~$\pm$~0.02; [$\alpha$/Fe]~=~+0.45~$\pm$~0.08. Thus, the stellar populations of the metal-rich globular clusters and underlying galaxy light, at a similar position in the galaxy, are similar to each other. This suggests a common formation history for the metal-rich globular clusters and the galaxy spheroid (see \citealt{brodie06}).

\section{CONCLUSIONS}
We find that the stellar population of NGC~1407 is uniformly old, 12.0~$\pm$~1.1~Gyr, with a supersolar degree of $\alpha$-elements enhancement, 0.41~$\pm$~0.05. We also measure a steep negative metallicity gradient of $-$0.38~$\pm$~0.04 and a central metallicity of 0.29~$\pm$0.08. The stellar population of NGC~1400 is found to be constantly old, 13.8~$\pm$~1.1 Gyr, and largely enriched by $\alpha$-elements, 0.33~$\pm$~0.04. The central metallicity is 0.25~$\pm$~0.06 and it steeply declines towards outer radii with a gradient of $-$0.47~$\pm$~0.04. 

The results of this work (and Paper~I) suggest a similar star formation and evolutionary history for the two galaxies. The outlined scenario is compatible with the revised version of the monolithic collapse model of galaxy formation and evolution. The two galaxies formed the bulk of their stars at z~$\geq$~5. The star forming episode lasted for no longer than 1~Gyr with a star formation rate $\geq$~100~M$_{\sun}$/year. NGC~1407 might have experienced an ancient merger, as inferred by the possible kinematically decoupled core (Paper~I); nevertheless, the detection is uncertain and potentially originated by a misalignment of the slit with respect to the centre of the galaxy, as suggested by the $h_{3}$ radial profile. In Paper~I we have also claimed that NGC~1400 has not interacted with NGC~1407 or the group intergalactic medium. The flatness of the $\alpha$-element radial profiles and the strong negative metallicity gradients are consistent with an outside-in formation scenario, where supernova-driven galactic winds played a fundamental role and shaped the slope of the gradients. The uniform radial profile of old ages, steep metallicity gradients and lack of signature of recent mergers indicate that the galaxies formed stars quickly some $\sim$~12~Gry ago and then have evolved quiescently without any significant interaction ever since.

\section{ACKNOWLEDGEMENTS}
Based on observations made with the NASA/ESA Hubble Space Telescope, obtained from the Data Archive at the Space Telescope Science Institute, which is operated by the Association of Universities for Research in Astronomy, Inc., under NASA contract NAS 5-26555. These observations are associated with program ID=9427 PI=Harris. We thank Lee Spitler for providing the Suprime-Cam photometric data. We also thank Fatma Reda for useful discussions on stellar population analysis. DF, RP and SB thank the ARC for financial support.

\begin{appendix}

\section{Calibrations to the Lick/IDS system}
\label{lickcalib}
The five following Lick/IDS and spectrophotometric standard stars were observed: HD4628 (K2), HD4656 (K5), HD69267 (K4), HD83618 (K2.5), HD97907 (K3).

After the correction for recession velocity redshift, the standard stars spectra were broadened to the Lick/IDS resolution. The procedure creates a resultant spectrum of total broadening:
\begin{equation}
\sigma_{G}^{2}= \sigma_{V}^{2}+\sigma_{L}^{2} \; , 
\end{equation}
where $\sigma_{V}^{2}$ is the velocity dispersion and $\sigma_{L}^{2}$ the appropriate Lick/IDS calibration resolution. The Lick/IDS spectral resolution for each index is estimated from Fig.~7 of \cite{wott97}, as in Table~2 of \cite{proctor02}. The Lick/IDS calibration resolution: 
\begin{equation}
\sigma_{L}^{2}= \sigma_{I}^{2}+\sigma_{B}^{2} \; ,
\end{equation}
is reached taking into account the instrumental broadening of our data $\sigma_{I}^{2}$ and convolving the spectrum with a Gaussian of width $\sigma_{B}^{2}$.
At this stage Lick/IDS indices were measured from the standard stars spectra and offsets between our index measurements and published values for the Lick/IDS standard stars estimated (Table~\ref{offset}).

\begin{table}
\begin{center}
\begin{tabular}{cccc}
\hline
\hline
Index & Unit & Lick/IDS Offset & Error in Mean\\
\hline
Fe4383  & \AA  & -0.229  &  0.120\\
Ca4455  &  \AA &-0.054  &   0.171\\
Fe4531  & \AA  &-0.342   &  0.103\\
C4668   & \AA  &-0.598   &  0.187\\
H$_{\beta}$& \AA &-0.041 &    0.113\\
Fe5015  & \AA & -0.357  &   0.185\\
Mg$_{1}$ & mag& 0.019   &  0.007\\
Mg$_{2}$ & mag& 0.018   &  0.006\\
Mg$_{b}$ & \AA& 0.039  &   0.195\\
Fe5270   & \AA& -0.412  &   0.147\\
Fe5335  & \AA& -0.274   &  0.215\\
Fe5406 &  \AA& -0.302  &   0.059\\
Fe5709 & \AA&  0.031  &  0.065\\
Fe5782 & \AA&  0.026   &  0.067\\
NaD    & \AA& -0.239  &   0.138\\
TiO$_{1}$ & mag& 0.006   &  0.003\\
\hline 
\hline
\end{tabular}
\end{center}
\caption{Offsets between our index measurements and published values of the five Lick/IDS and spectrophotometric standard stars.}
\label{offset}
\end{table}

The Fourier quotient technique within IRAF software package was used to estimate recession velocity and velocity dispersion for each aperture spectrum. The kinematic measurements are in good agreement with those obtained with van der Marel's technique (Paper~I). We calibrated adopting the same technique for the standard stars. The method is applicable when the measured velocity dispersion satisfies: 
\begin{equation}
\sigma_{V}^{2} + \sigma_{I}^{2} \leq \sigma_{L} \; .
\end{equation}
\noindent When the opposite condition is found, the aperture spectra are left un-broadened. Lick/IDS indices were then measured and index correction factors applied. 

Th index correction factors were estimated using the method described in \cite{proctor02} (Appendices A1 and A2). Briefly, the Lick/IDS indices were measured in the 5 Lick/IDS standard stars after convolving the spectra with a series of Gaussians of known widths. Consequently, the appropriate correction factor for a specific index $i$ was obtained by a polynomial fit of order~3
\begin{equation}
C_{i}= x_{0} + x_{1}\sigma_{C} + x_{2}\sigma_{C}^{2} + x_{3}\sigma_{C}^{3} \; ,
\end{equation}
\noindent where $\sigma_{C} = \sqrt{\sigma_{I}^{2} + \sigma_{V}^{2} - \sigma_{L}^{2}}$, to the correction factors of the index $i$ (average of the values from the 5 standard stars), at each of the Gaussian broadenings.

The offsets evaluated from the Lick/IDS library stars were applied to the Lick/IDS indices measured from the 1D galaxy spectra. Index errors arise from two types of errors which are added in quadrature: i.e. random statistical errors due by Poisson noise and uncertainties in recession velocity and velocity dispersion and systematic errors originated from the conversion onto the Lick/IDS system, quantified by the error on the mean of the offsets. Systematic and random errors are of the same order.

\section{Lick/IDS line-strength radial profiles}
\begin{table}
\begin{center}
\begin{tabular}{cccc}
\hline   
\hline
                 & NGC~1407 & & \\
\hline
Index            & $\frac{\Delta Index}{\Delta \log r}$ &  $Index_{r_{e}}$ & $Index_{0}$ \\
\hline
Fe4383  &       -0.65 $\pm$ 0.14 & 4.40 $\pm$ 0.16 & 5.21 $\pm$ 0.46  \\
Ca4455  &  	-0.27 $\pm$ 0.09 & 1.49 $\pm$ 0.10 & 1.86 $\pm$ 0.29  \\
Fe4531  & 	-0.37 $\pm$ 0.10 & 2.77 $\pm$ 0.12 & 3.24 $\pm$ 0.35   \\
C4668   & 	-2.50 $\pm$ 0.16 & 4.55 $\pm$ 0.18 & 7.75 $\pm$ 0.53   \\
H$_{\beta}$&    -0.01 $\pm$ 0.07 & 1.28 $\pm$ 0.08 & 1.29 $\pm$ 0.22   \\
Fe5015  & 	-1.83 $\pm$ 0.14 & 3.44 $\pm$ 0.16 & 5.84 $\pm$ 0.48   \\
Mg$_{1}$ & 	-0.04 $\pm$ 0.00 & 0.12 $\pm$ 0.00 & 0.18 $\pm$ 0.01   \\
Mg$_{2}$ &	-0.08 $\pm$ 0.00 & 0.24 $\pm$ 0.00 & 0.34 $\pm$ 0.01  \\
Mg$_{b}$ &      -1.08 $\pm$ 0.09 & 3.71 $\pm$ 0.10 & 5.12 $\pm$ 0.30   \\
Fe5270  & 	-0.81 $\pm$ 0.08 & 1.64 $\pm$ 0.09 & 2.68 $\pm$ 0.27   \\
Fe5335  & 	-0.80 $\pm$ 0.11 & 1.49 $\pm$ 0.12 & 2.54 $\pm$ 0.36   \\
Fe5406  &  	-0.52 $\pm$ 0.06 & 0.83 $\pm$ 0.07 & 1.52 $\pm$ 0.22   \\
Fe5709  &       -0.20 $\pm$ 0.05 & 0.68 $\pm$ 0.05 & 0.94 $\pm$ 0.15   \\
Fe5782  & 	-0.31 $\pm$ 0.05 & 0.55 $\pm$ 0.05 & 0.96 $\pm$ 0.15  \\
NaD     & 	-2.05 $\pm$ 0.07 & 2.41 $\pm$ 0.08 & 5.07 $\pm$ 0.23   \\
TiO$_{1}$ &	-0.01 $\pm$ 0.00 & 0.04 $\pm$ 0.00 & 0.05 $\pm$ 0.00  \\
\hline 
\hline
\end{tabular}
\end{center}
\caption{Results of linear regression analysis to NGC~1407 Lick/IDS indices. Logarithmic radial gradients, values at the effective radius (measured by the intercept at $r=r_{e}$) and central index values (average of the values extracted in all the apertures within a radius of $r_{e}/8$) are given.}
\label{line1}
\end{table}

\begin{table}
\begin{center}
\begin{tabular}{ccccccc}
\hline   
\hline
                 & NGC~1400 & & \\
\hline
Index            &  $\frac{\Delta Index}{\Delta \log r}$ &  $Index_{r_{e}}$ & $Index_{0}$ \\
\hline
Fe4383  &        -1.16 $\pm$ 0.17 & 3.74 $\pm$ 0.14 & 5.10 $\pm$ 0.45  \\
Ca4455  &  	 -0.43 $\pm$ 0.10 & 1.22 $\pm$ 0.08 & 1.74 $\pm$ 0.29  \\
Fe4531  & 	 -0.80 $\pm$ 0.13 & 2.49 $\pm$ 0.11 & 3.46 $\pm$ 0.34  \\
C4668   & 	 -2.82 $\pm$ 0.19 & 4.19 $\pm$ 0.16 & 7.47 $\pm$ 0.52  \\
H$_{\beta}$&      0.11 $\pm$ 0.08 & 1.39 $\pm$ 0.07 & 1.26 $\pm$ 0.21  \\
Fe5015  & 	 -1.65 $\pm$ 0.17 & 3.79 $\pm$ 0.14 & 5.63 $\pm$ 0.47  \\
Mg$_{1}$ & 	 -0.06 $\pm$ 0.00 & 0.11 $\pm$ 0.00 & 0.17 $\pm$ 0.01  \\
Mg$_{2}$ &	 -0.10 $\pm$ 0.00 & 0.21 $\pm$ 0.00 & 0.33 $\pm$ 0.01  \\
Mg$_{b}$ &       -1.69 $\pm$ 0.10 & 3.43 $\pm$ 0.08 & 5.36 $\pm$ 0.30  \\
Fe5270  & 	 -0.85 $\pm$ 0.09 & 1.85 $\pm$ 0.08 & 2.81 $\pm$ 0.26  \\
Fe5335  & 	 -1.16 $\pm$ 0.12 & 1.72 $\pm$ 0.10 & 2.94 $\pm$ 0.36  \\
Fe5406  &  	 -0.62 $\pm$ 0.08 & 1.17 $\pm$ 0.06 & 1.90 $\pm$ 0.22  \\
Fe5709  &        -0.07 $\pm$ 0.05 & 0.83 $\pm$ 0.05 & 0.91 $\pm$ 0.15  \\
Fe5782  & 	 -0.43 $\pm$ 0.05 & 0.58 $\pm$ 0.05 & 1.01 $\pm$ 0.15  \\
NaD     & 	 -2.52 $\pm$ 0.08 & 2.49 $\pm$ 0.06 & 5.37 $\pm$ 0.22  \\
TiO$_{1}$ &	 -0.01 $\pm$ 0.00 & 0.03 $\pm$ 0.00 & 0.05 $\pm$ 0.00  \\
\hline 
\hline
\end{tabular}
\end{center}
\caption{Results of linear regression analysis to NGC~1400 Lick/IDS indices. Logarithmic radial gradients, values at the effective radius (measured by the intercept at $r=r_{e}$) and central index values (average of the values extracted in all the apertures within a radius of $r_{e}/8$) are given.}
\label{line2}
\end{table}

\label{lsf}
Logarithmic gradients of the indices were measured by applying a linear regression analysis to the measured index values:
\begin{equation}
Index(r) = Index_{r_{e}} + \frac{\Delta Index}{\Delta \log r} \log\frac{r}{r_{e}} \; ,
\end{equation}
where $r$ is the galactocentric radius, $r_{e}$ is the effective radius of the galaxy and $Index_{r_{e}}$ is the index value defined at the effective radius. We accounted for observational errors introducing the weights $\omega_{i} = \sigma_{i}^{-2}$, where $\sigma_{i}$ are the errors associated to measured values, $y_{i}$, of a specific Lick/IDS index at the position $x_{i}$ ($\equiv \log(r_{i}/r_{e})$). The former equation is then equivalent to $y = A +Bx$, where
\begin{equation}
A= (\sum^{n}_{i=1} \omega_{i} x_{i}^{2} \sum^{n}_{i=1} \omega_{i} y_{i} - \sum^{n}_{i=1} \omega_{i} x_{i} \sum^{n}_{i=1} \omega_{i} x_{i} y_{i}) \Delta^{-1}
\end{equation}
and
\begin{equation}
B=(\sum^{n}_{i=1} \omega_{i} \sum^{n}_{i=1} \omega_{i} x_{i}y_{i} - \sum^{n}_{i=1} \omega_{i} x_{i} \sum^{n}_{i=1} \omega_{i} y_{i}) \Delta^{-1}
\end{equation}
where 
\begin{equation}
\Delta = \sum^{n}_{i=1} \omega_{i} \sum^{n}_{i=1} \omega_{i} x_{i}^{2} - (\sum^{n}_{i=1} \omega_{i} x_{i})^{2} \; .
\end{equation}
The uncertainties on $A$ and $B$ are:
\begin{equation}
\sigma_{A} = \sqrt{(\sum^{n}_{i=1} \omega_{i} x_{i}^{2})\Delta^{-1}} \; , 
\qquad \sigma_{B} = \sqrt{(\sum^{n}_{i=1} \omega_{i})\Delta^{-1}} \; .
\end{equation}

Tables ~\ref{line1} (NGC~1407) and \ref{line2} (NGC~1400) report the results of the linear regression analysis. We derived logarithmic radial gradients for all indices and the indices values at the effective radius. Central index values were estimated by averaging the values extracted in all the apertures within a radius of $r_{e}/8$.

\begin{figure*}
\includegraphics[scale=.8]{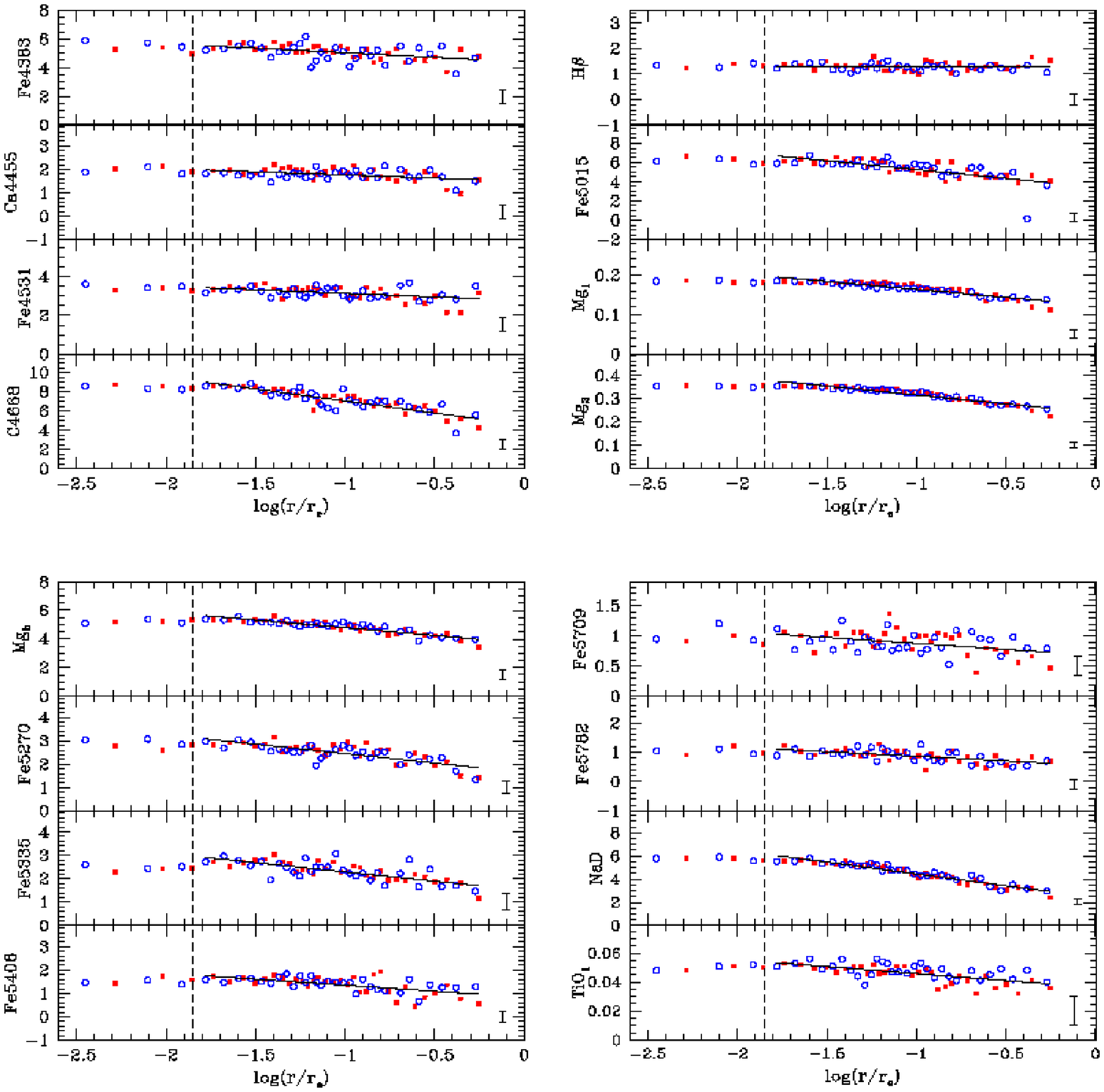}
\caption{Spatially resolved NGC~1407 Lick/IDS line-strength indices radial profiles. Filled red squares represent values from the aperture spectra extracted along the observing axis in the negative binning direction; open circles are the same in the positive binning direction (see text for details). The approximate extent of the seeing disc is identified by a dashed line. The error bars express the median value in the errors data set in the derived line-strength indices. Gradients (black thick lines) are measured excluding the points inside the region affected by the seeing. The quantities are measured in logarithmic radius scaled with the effective radius of the galaxy for an easy comparison to literature data.}
\label{n1407multirad}
\end{figure*}

\begin{figure*}
\includegraphics[scale=.8]{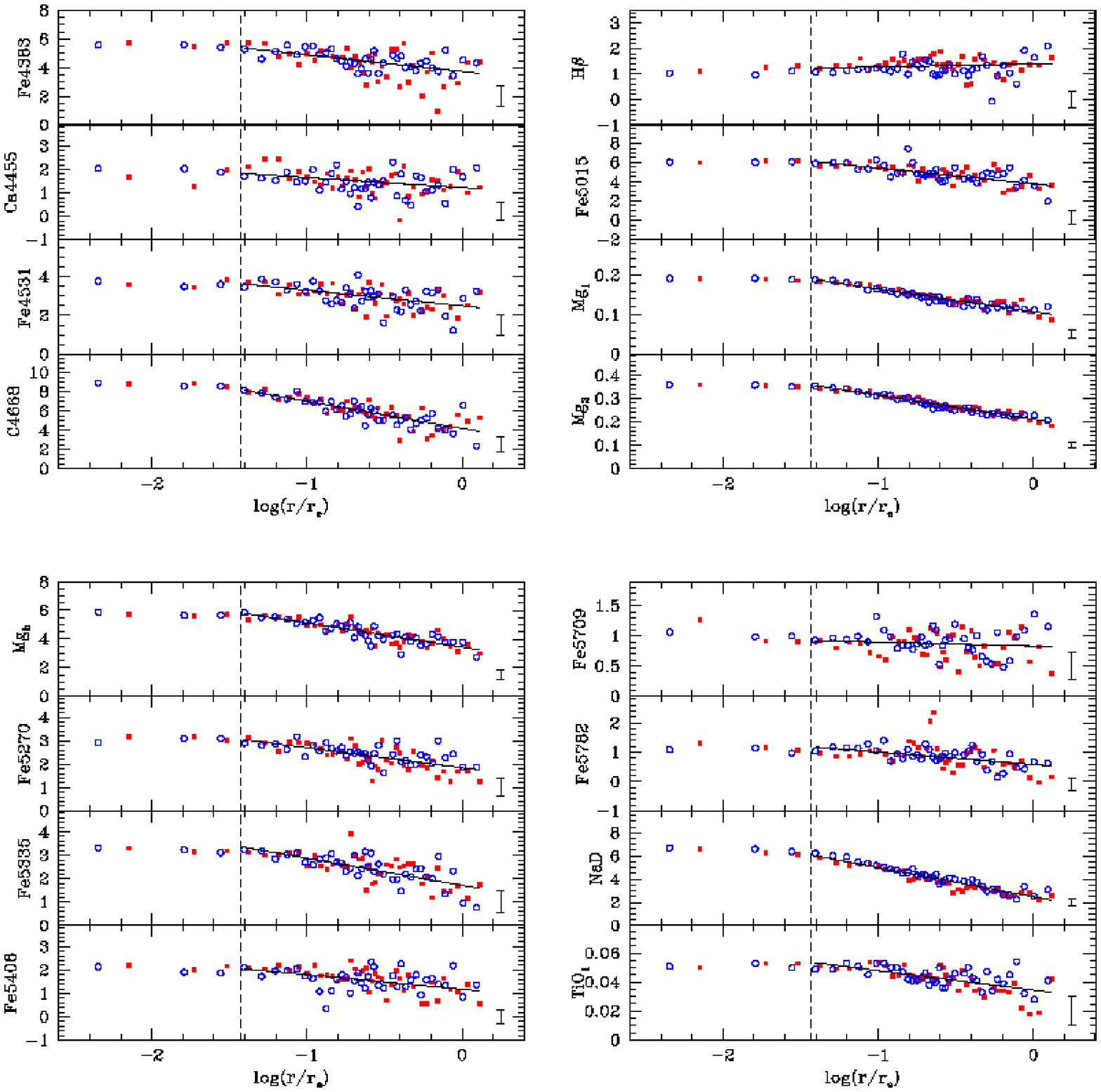}
\caption{Spatially resolved NGC~1400 Lick/IDS line-strength indices radial profiles. Symbols and description are as in Fig.~\ref{n1407multirad}.}
\label{n1400multirad}
\end{figure*}
\end{appendix}
\end{document}